\documentclass{aa}  
 
\bibpunct{(}{)}{;}{a}{}{,}

\usepackage{graphicx}
\usepackage{txfonts}
\newcommand{\teff}{\ensuremath{T_{\rm eff}}}
\newcommand{\bs}{\ensuremath{\langle \vert B \vert \rangle}}
\newcommand{\bz}{\ensuremath{\langle B_z \rangle}}
\begin{document} 

   \title{Searching for magnetic fields in featureless white dwarfs\\ with the DIPOL-UF polarimeter at the Nordic Optical Telescope}

   \titlerunning{Searching for magnetic fields in featureless white dwarfs\\ with DIPOL-UF at NOT}

   \author{A. Berdyugin\inst{1}
          \and
          J. D. Landstreet\inst{2,3}
          \and
          S. Bagnulo\inst{2}
          \and
          V. Piirola\inst{1}
          \and
          S. V. Berdyugina\inst{4,5,6}
          }

   \institute{Department of Physics and Astronomy, FI-20014 University of Turku, Finland\\
              \email{andber@utu.fi%, piirola@utu.fi
              }
         \and
            Armagh Observatory \& Planetarium, College Hill, Armagh BT61 9DG, UK\\
             \email{%John.Landstreet@Armagh.ac.uk,
         Stefano.Bagnulo@Armagh.ac.uk}
             \and
            University of Western Ontario, London, Ontario N6A 3K7, Canada \\
             \email{jlandstr@uwo.ca}
            \and
            IRSOL Istituto Ricerche Solari “Aldo e Cele Dacc\`o", Faculty of Informatics, Universit\`a della Svizzera italiana, Via Patocchi 57, Locarno, Switzerland%\\
            %\email{svetlana.berdyugina@irsol.usi.ch}
            \and
            Euler Institute, Faculty of Informatics, Universit\`a della Svizzera italiana, Via la Santa 1, 6962 Lugano, Switzerland
            \and
            Institut f\"ur Sonnenphysik (KIS), Georges-K\"ohler-Allee 401A, 79110 Freiburg, Germany\\
%%%             \email{e@e}
             }

   \date{Received April 1, 2024; accepted }

% \abstract{}{}{}{}{} 
% 5 {} token are mandatory
\abstract{About 20\% of the white dwarfs possess a magnetic field that may be detected by the splitting and/or polarization of their spectral lines. As they cool, the effective temperatures of the white dwarfs becomes so low that no spectral lines can be seen in the visible wavelength range. If their atmospheres are not polluted by the debris of a planetary system, these cool white dwarfs have featureless optical spectra. Until quite recently, very little was known about the incidence of magnetic fields in these objects. However, when observed with polarimetric techniques, a significant number of featureless white dwarfs reveal  strong magnetic fields in their optical continuum spectra. Measuring the occurrence rate and strength of magnetic fields in old white dwarfs may help us to understand how these fields are generated and evolve. We report the results of an ongoing survey of cool white dwarfs with the high-precision broad-band polarimeter DIPOL-UF, which is deployed at the Nordic Optical Telescope on La Palma, Spain. This survey has led to the firm discovery of 13 cool magnetic white dwarfs in the solar neighborhood so far, including six new detections that we report in this paper. 
}
  \keywords{polarization -- stars: white dwarfs -- stars: magnetic fields}

   \maketitle
%
%-------------------------------------------------------------------

%%%%%%%%%%%%%%%%%%%%%%
\section{Introduction}
%%%%%%%%%%%%%%%%%%%%%%
White dwarfs (WDs) are the final evolution stage of at least 90\% of all stars when the useable nuclear fuel is exhausted. These objects ($M \sim 0.6 M_\odot$, $R \sim 0.01 R_\odot$) are common in space around the Sun; about 150 of them are found within a distance of 20\,pc \citep{Holletal18}. The basic evolution of a WD, which initially formed by the collapse of a stellar core, is to cool slowly on a timescale of several billion years. {The physics of WDs has recently been reviewed by \citet{Sauetal22}. } However, this evolution is found to be surprisingly complex, for example because magnetic fields are present in more than 20\,\% of the WDs \citep[e.g.,][]{BagLan21}.

Magnetic fields have been detected in several hundred nearby WDs, mostly through Zeeman splitting of spectral lines that were observed in flux spectra \citep[e.g.][]{Feretal20}. The fields cover a wide range of strengths, from below $10^4$\,G up to nearly $10^9$\,G. In a volume-limited survey within 20\,pc around the Sun, fields were detected in 20--25\% of the WDs that were surveyed \citep{BagLan21}. The fields do not appear to evolve with time on an observable timescale, although they are often not symmetric about the rotation axis of the host WD and thus appear to vary periodically as the star rotates. {A geometrical model for a star with a dipolar field that is tilted with respect to the rotation axis was developed for the magnetic Ap and Bp stars by \citet{Stibbs50}, and it is still used to describe many magnetic WDs \citep[see, e.g.,][]{Bagetal24}. } In this respect, magnetic WDs (MWDs) are rather similar to the small fraction ($\sim 7$\,\%) of magnetic upper main-sequence stars \citep[magnetic Ap, Bp and O stars: see, e.g.,][]{DonLan09}, which also exhibit stable magnetic fields with a typical strength of about 1-100\,kG. These intrinsically stable fields are referred to as "fossil fields". That these fields exist in only a fraction of WDs and of upper main-sequence stars, that is, in settings without an obvious dynamo activity, is one of the outstanding physical puzzles of stellar evolution. 

Studies of volume-limited samples of MWDs have revealed two specific evolution paths. (1) WDs resulting from single-star evolution, with typical masses of about $0.6 M_\odot$, only very rarely reveal detectable fields during at least the first billion years of cooling, and the fields observed in these young WDs are almost always very weak, typically tens or hundreds of kiloGauss (kG). As the WDs cool, after about 2--3\,Gyr, the fields begin to appear much more frequently and with much greater field strength \citep{BagLan22}. (2) The most massive WDs, with masses above perhaps $1.1\,M_\odot$, are probably produced by WD-WD binary mergers {\citep{Touetal08,Brietal15,Garetal12}}. These massive WDs are usually strongly magnetic from nearly (or exactly) the moment of the merger, with fields of many megaGauss (MG) and very short rotation periods \citep{BagLan22}. Intermediate-mass WDs probably represent a mixed population with a variety of origins, and may frequently result from binary evolution.

However, our observational understanding of the later evolution of magnetic fields in WDs that are older than a cooling age of about 5\,Gyr is extremely limited. By this age, most magnetic WDs no longer display detectable Balmer lines (He lines disappear at a much younger age), which are commonly employed for detecting a magnetic field through spectral lines. If, however, WDs are metal polluted from accretion of circumstellar debris (often classified as DZ WDs), metal lines can be employed {\citep{Holetal17,Holetal18,Kawetal19}}. Alternatively, the magnetism of cool carbon-rich WDs with strong carbon-based molecular bands, classified as DQ WDs, can be studied using polarization in CH molecular bands when they are detectable {e.g.,\citet{AngLan72}}, \citet{Berdetal07}, and \citet{Voretal13}. In the remaining featureless cool WDs, which are classified as DC WDs, a broad-band polarimetry may be employed. This technique has hardly been used for magnetic field searches since the 1970s \citep{Angetal81}. After these earliest works, the only discoveries of magnetic DC WDs were those made with spectropolarimetry by \citet{Putney97} and by \citet{BagLan20}. Before the survey of magnetic WDs in the local 20\,pc volume was completed \citep{BagLan21}, only 10 of the 30 DC WDs in this volume (the nearest and brightest DC WDs) had been searched for evidence of magnetism, and virtually none of the fainter, more distant DCs had been studied. This lack of data meant that we had almost no information about the presence of magnetic fields in the commonest spectral class of the oldest WDs. Hence, the magnetism in WDs that was produced during the first half of the age of the disk of the Milky Way was almost completely unknown territory. 

In order to fill some of this information void, we are carrying out a systematic survey of optical circular polarization of known and suspected DC WDs in the northern hemisphere \citep{Beretal22,Beretal23}. Here, we present a compilation of more than 110 new observations obtained for 84 WDs in 2022 July, 2023 November, and 2024 February with the high-precision three-band polarimeter DIPol-UF (Double Image Polarimeter - Ultra Fast) deployed at the Nordic Optical Telescope (NOT), located at the Observatorio del Roque de los Muchachos on the Canary Island of La Palma, Spain, including six new detections of magnetic DC WDs.

%%%%%%%%%%%%%%%%%%
\section{Observations}
%%%%%%%%%%%%%%%%%%
New broad-band circular polarization (BBCP) measurements were obtained from 2023 March -- 2024 February using the DIPol-UF three-band polarimeter on the 2.5 m NOT. This instrument has been described in detail by \citet{Piietal21}. Briefly, we obtain measurements of broad-band circular polarization in three bands with DIPol-UF (which we call {\it B', V', R'}), each about 1000\,\AA\ wide, centered on the wavelengths 4450\,\AA, 5400\,\AA, and 6400\,\AA. These bands  are defined by sharp-cutoff dichroic filters with a transmission of about 90\,\% across most of each band \citep[see Fig.~1 of][]{Beretal22}. 
The stability, sensitivity, and zeropoint errors of DIPol-UF have been studied extensively using observations of bright standard stars and WDs with known circular polarization properties, and it is clear that we can reliably detect polarization levels down to $0.01$\,\%\ with a precision and accuracy of 0.002\,\% at least. In practice, the precision of the survey is mainly limited by the photon number  from each target WD at a given telescope \citep{Piietal21,Beretal22}. 

\subsection{Observations of standard stars}
During our observing run, we obtained measurements of various bright nonpolarized nearby stars to calibrate the instrumental polarization. 
In addition, we observed the well-known MWD WD\,1900+705, which appears to show a signal of circular polarization that is nearly constant with time \citep[see, e.g.][]{BagLan19a}. 

The measurements of unpolarized stars yield an instrumental polarization to a precision better than $10^{-5}$ (a few parts per million). In the $B'V'R'$ bands, the values of the reduced Stokes parameter $V/I$ are 0.0121 $\pm$ 0.0004 \%, 0.0109 $\pm$ 0.0005 \%, and 0.0084 $\pm$ 0.0004 \%, respectively.  This instrumental polarization was subtracted from the measured polarization of all targets.

\subsection{Science targets}
Most DC WDs have relatively low effective temperature $T_{\rm eff}$ values, often lower than 5000\,K, and the nearest DC WDs mostly have \textit{Gaia G} magnitudes of 15 or fainter. All our targets were chosen based on the recent nearly complete survey of spectral types and physical parameters of northern WDs within 40\,pc of the Sun \citep{McCetal20,OBretal24}. By expanding our survey to a volume within roughly 40\,pc of the Sun, we reach typical magnitudes of 17 or even fainter for the most interesting (oldest) DCs. This is near or beyond our limiting magnitude for the desired measurement uncertainty level.

%%%%%%%%%%%%%%%%
%%%\input{Table_Program}
%%%%%%%%%%%%%%%%

Table~\ref{Table_Program} gives the list of DC WDs that we observed during four observing runs on the NOT in 2022 November, 2023 March and November, and 2024 February. We also included some observations that were obtained in a previous run in July 2022, which were not published in our previous papers because of an oversight. The target names are listed according to the (frequently unofficial but convenient) Villanova designation  \citep{McCSio77,McCSio99}, followed by the main {\it Simbad} identifier, the \textit{Gaia G} magnitude and distance, and various stellar parameters derived from \citet{Bloetal19} and \citet{Genetal21}. The last column of Table~\ref{Table_Program} shows the spectral class of the WD when it is different than DC, using the classification by \citet{OBretal24} (see below).
In addition to targets that were never observed before, our list includes 
a number of WDs in which a magnetic field has been detected in previous studies, namely
WD\,0004$+$122 \citep{BagLan20};  %1
WD\,0548$-$001 \citep[spectral class: DQp][]{LanAng71};  %2
WD\,0756$+$437 \citep{Putney97};  %3
WD\,1008$+$290 \citep[spectral class: DQ,][]{Schetal99}; %4
WD\,1036$-$204 \citep[spectral class: DQ,][]{Lieetal78}; %5
WD\,1315$+$222 \citep{Beretal22}; %6
WD\,1346$+$121 \citep{Beretal23}; %7
WD\,1556$+$044 \citep{Beretal22}; %8
WD\,2049$-$222 \citep{Beretal22}; %9
WD\,2049$-$253 \citep{BagLan20};  %10
WD\,2211$+$372 \citep{Beretal23}. %11
These observations served to confirm detection, to further verify the instrument reliability and stability, and to set a baseline for future studies of the rotational periods of these MWDs.

This target list includes different categories of cool WDs: He-rich and H-rich featureless WDs, and a small number of cool DQ WDs with molecular carbon (C$_2$) bands. WDs with He-rich atmospheres cease to display spectral lines when the cooling has reduced \teff\ below about 11\,000\,K, which occurs at a cooling age of about 1\,Gyr, and our list includes a number of DC WDs of this type that have ages of only 2--4\,Gyr.  In contrast, the Balmer lines of WDs with H-rich atmospheres are still visible until \teff\ has dropped to a value of about 5000\,K, which requires a cooling time of about 5\,Gyr. 

%%%%%%%%%%%%%
%%%\longtab{
%%%\input{Table_WD2_Log}
%%%}
%%%%%%%%%%%%

The resulting measurements of the broad-band circular polarization in each of the three filter bands, together with timing information, including JD, date, and time at the middle of each observation, and the total exposure time of each target, are listed in Table\,\ref{Tab_WD_Log}. 
Each observation consisted of a long series of 30s to 60s exposures. We report in Table\,\ref{Tab_WD_Log} the average of these series values, hereafter called measurements, while the mean quadratic error from the individual exposures of a given series was used to estimate the measurement error. Additional details of the data acquisition and reduction are discussed in \citet{Beretal22}. 
%%%%%%%%%%%%%%%%%%%%%%%%%%%%%%%%%%%%%%%%%%%%%%%%%%%%%%%%%%%%%%%%%%%%%%%%%%%%%%%%%%%%%%%%%%%%%%%%%%%%%%%%%%%%%%%%%%%%%%%%%%%%%
\begin{figure}[ht]
\centering
\includegraphics[width=15.2cm,trim={2.0cm 3.4cm 4.3cm 1.0cm},clip]{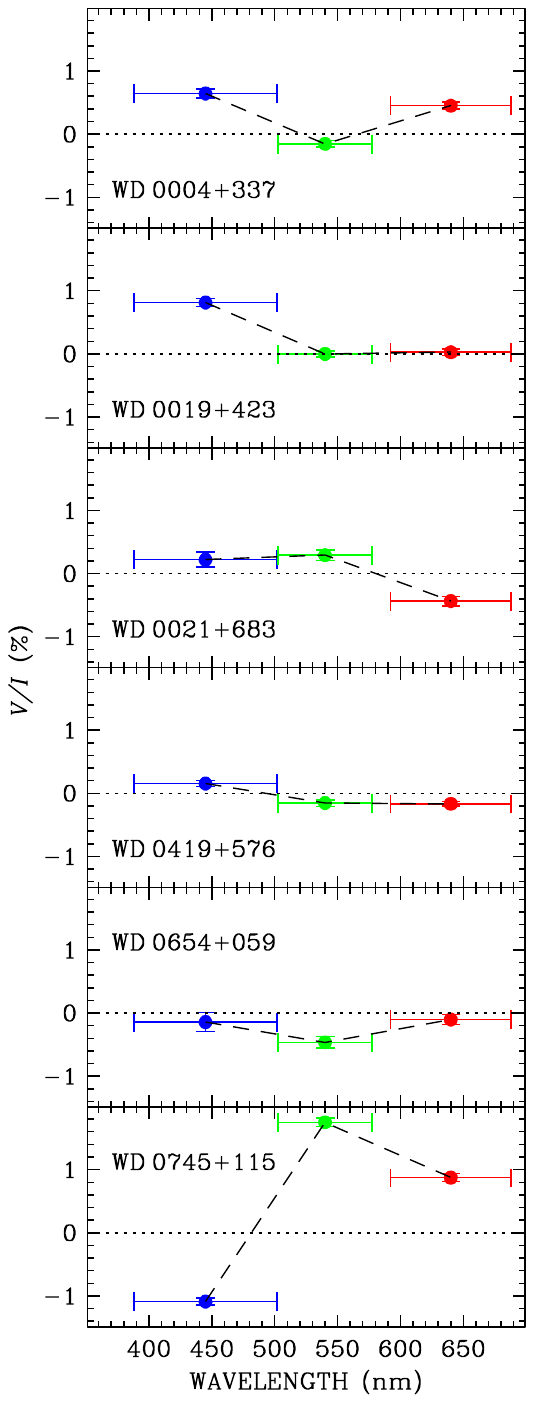}
      \caption{\label{Fig_NOT} Wavelength dependence of the circular polarization for the six newly detected magnetic WDs.} 
\end{figure}
%%%%%%%%%%%%%%%%%%%%%%%%%%%%%%%%%%%%%%%%%%%%%%%%%%%%%%%%%%%%%%%%%%%%%%%%%%%%%%%%%%%%%%%%%%%%%%%%%%%%%%%%%%%%%%%%%%%%%%%%%%%%%

Inspection of Table\,\ref{Tab_WD_Log} shows that for the hotter targets, the uncertainties tend to be similar across all three bands, and they tend to be smallest in the red $R'$ band for the cooler WDs. For obvious reasons, the uncertainties are larger for the fainter WDs, for which they sometimes rise above 0.1\,\%. Single-filter measurements that show a detection of a nonzero circular polarization at the $3\sigma$ level or better are indicated in Table\,\ref{Tab_WD_Log} by  boldface fonts. We considered that we detected nonzero circular polarization either (1) when more than one measurement in any filter band exceeded a significance of $3\sigma$, or (2) when a single measurement was obtained with a significance of $5\sigma$. With these criteria, we detected fields in 6 of the 75 new WDs reported in this study. These stars are indicated by boldface fonts for their names in Tables\,\ref{Table_Program} and \ref{Tab_WD_Log}. Two of the six fields we detected were further confirmed by the detection of similar polarization levels in two successive measurements. The names of the new magnetic WDs discovered in this survey are highlighted with an asterisk in Tab.~\ref{Tab_WD_Log}.

%%%%%%%%%%%%%%%%%%%%%%%%%%%%%%%%%%%%%%%%%%%%%%%%%%%%%%%%%%%%%%%%%%%%%%%%%%%%%%%%%%%%%%%%%%%%%%%%%%%%%%%%%%%%%%%%%%%%%%%%%%%%%
\begin{figure}
\centering
\includegraphics[width=8.7cm,trim={0.7cm 5.8cm 1.1cm 3.0cm},clip]{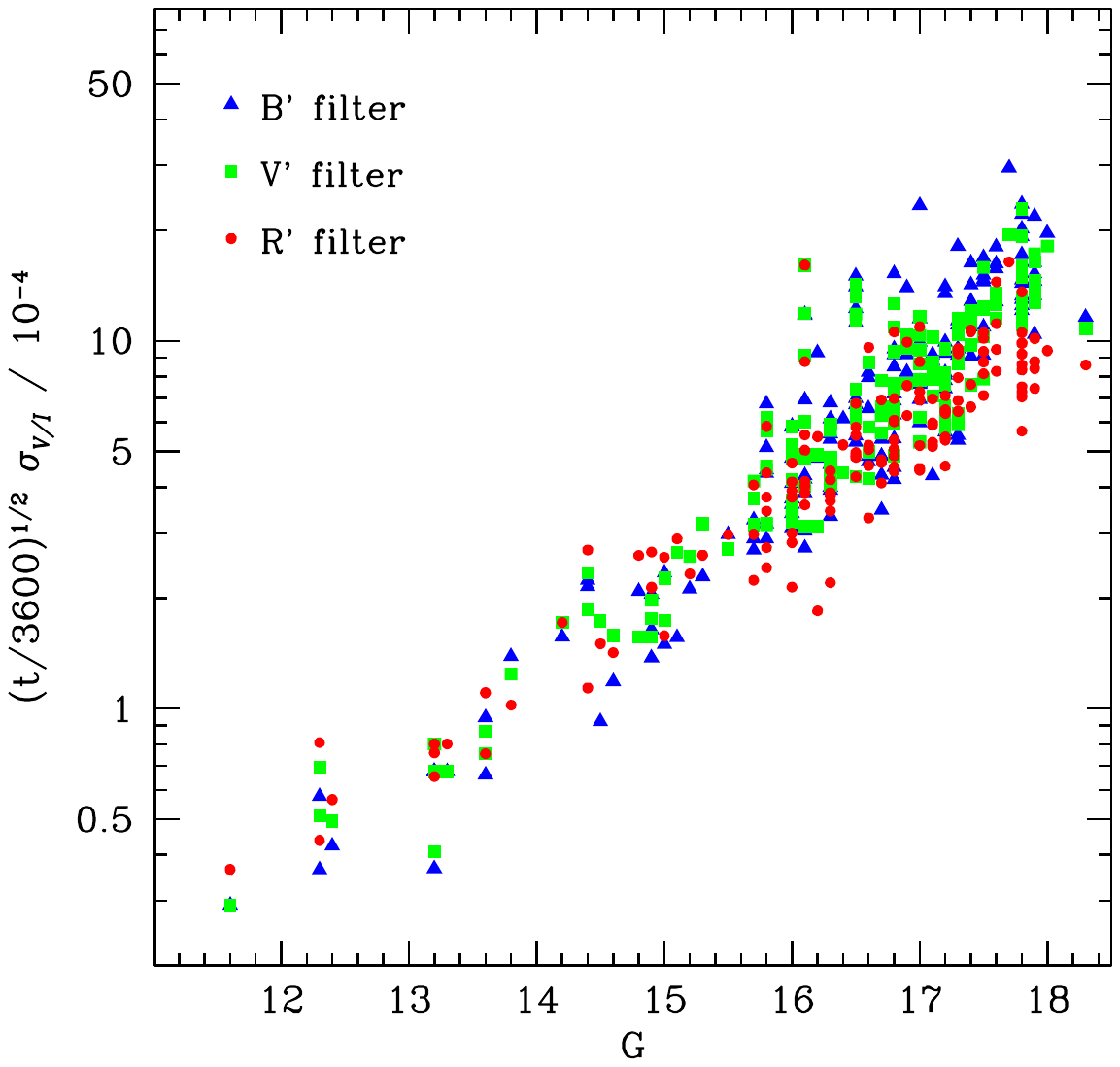}
      \caption{
      \label{Fig_ETC} 
      Relation of the magnitude and measurement errors of all DIPoL-UF measurements of WDs at NOT as a function of \textit{Gaia} $G$ magnitude, normalized to 1\,h exposure time.
      } 
\end{figure}
%%%%%%%%%%%%%%%%%%%%%%%%%%%%%%%%%%%%%%%%%%%%%%%%%%%%%%%%%%%%%%%%%%%%%%%%%%%%%%%%%%%%%%%%%%%%%%%%%%%%%%%%%%%%%%%%%%%%%%%%%%%%%

\section{Results}
\subsection{Newly discovered magnetic WDs}
Among the stars that were observed for the first time in this survey, we detected a nonzero signal of circular polarization in six WDs:
%out of 72 that had never been observed before with polarimetric techniques:
WD\,0004$+$337 = LP 240-29;  WD\,0019$+$423 = EGGR 459; WD\,0021$+$683 = G 242-54B; WD\,0419$+$576 = LP 84-16;  WD\,06542$+$059 = 2MASS J06572938+0550479; and WD\,0745$+$115 = SDSS~J074842.48+112502.0. Their polarization as a function of the broad-band filters is shown in Fig.~\ref{Fig_NOT}. 
 We comment on the newly discovered stars below. In addition, we comment on the known very strong field MWD WD\,0745$+$115, which shows short-term variability.

\subsubsection{WD\,0004$+$337}

This WD was originally identifed by Luyten (1995) as a nearby object based on its high proper motion. Its was classified as a probable DC WD by \citet{Limetal15} and was confirmed using \textit{Gaia} data by \citet{Genetal19,Genetal21}. Spectral class DC and physical parameters were determined by \citet{Bloetal19}. Most recently, the physical parameters have been slightly revised by \citet{OBretal24}. Because \teff is significantly above 5000\,K without any trace of H$\alpha$, it is assumed to have a He-rich atmosphere. The cooling age of this DCH star is estimated to be nearly 5\,Gyr (the letter H after the spectral type DC means that the WD is magnetic). 

We only observed this WD once, but nonzero polarization at about the 0.5\,\% level was observed in the {\it B'} filter at the $9 \sigma$ level and in the {\it R'} filter at the $8 \sigma$ level. We therefore regard this detection as quite secure. The polarization detected in both filters has the same sign. However, remarkably, a barely detectable polarization level (at 3$\sigma$) was observed in the {\it V'} filter.

\subsubsection{WD\,0019$+$423}

This object, first identified as a WD by \citet{Greenstein1979}, has been included in numerous studies. Its parameters  have most recently been derived by \citet{OBretal24}, who considered it to have a DC spectrum with a He-rich atmosphere. However, with a derived mass of only $0.39 M_\odot$, which is slightly below the lowest mass that can be produced by single-star evolution, we suspect that this object may be an undetected binary pair of two similar WDs.

Circular polarization is detected at roughly the 0.8\,\% level only in the {\it B'} band, but the observed values are nonzero at the 12 $\sigma$ levels, and the detection is therefore very secure. If our suspicion of binarity is correct and roughly half the light is contributed by a nonpolarized WD, the typical polarization levels in the two bands showing clear nonzero polarization could be near the 1.6\,\% level. In any case, the broad-band variation of the polarization is rather similar to that of WD\,0004$+$337 above. 

\subsubsection{WD\,0021$+$683}

This WD is the fainter member of a common-proper-motion binary with a separation of about 5\,arcsec. The companion is a very cool star some three mag brighter than the WD, presumably an M dwarf. The object was identified as a WD by \citet{Genetal19} and was observed spectroscopically by \citet{Treetal20}. It is included in the 40 pc WD sample \citep{McCetal20,OBretal24}. It appears to have a He-rich atmosphere and an age of more that 5\,Gyr. 

Circular polarization is only detected with high confidence in the {\it R'} band, where the polarization of 0.435\% provides a detection at nearly the $6 \sigma$ level. Polarization is also detected in the {\it V'} band at the $3.5 \sigma$ level, so that the field in this star is considered to be reliably detected, but is close to the threshold for a reliable detection with our telescope and instrument. 

This WD shows only weak polarization in the {\it B'} band and appears to have a BBCP spectrum that changes sign at a wavelength between the {\it R'} band and the two bluer bands. 

\subsubsection{WD\,0419$+$576}

This object is a relatively warm DC WD with a cooling age of about 2.5\,Gyr. It was identified as a WD candidate from the \textit{Gaia} DR2 data \citep{Jimenez-Esteban2018}. A spectrum obtained by \citet{Treetal20} led to a clear identification as a DC WD. The WD is within the 40\,pc volume \citep{McCetal20,OBretal24}.

The absolute value of the observed circular polarization signal is consistently about 0.15\%\ in the three bands, but it changes sign between the {\it B'} and the other two bands. In the {\it B'} and {\it V'} bands, the signal is only detected at about the $3 \sigma$ level, but in the {\it R'} band, the detection is at the $5 \sigma$  level, and we consider that the detection of a field is secure, if uncomfortably close to our detection limit. The magnetic field of this DCH star is probably somewhat weaker than those of the three WDs discussed above.

\subsubsection{WD\,0654$+$059}

This object is a recently discovered nearby WD. It was first identified in the \textit{Gaia} DR2 release by \citet{Jimenez-Esteban2018}. The WD is located within the local 40 pc volume, and the physical parameters were estimated by \citet{Genetal19,Genetal21,McCetal20,OBretal24}. A flux spectrum was obtained by \citet{Treetal20}, which resulted in the DC classification. 

Three band polarization measurements were obtained during two winter runs. Polarization was clearly detected first in one band and then in two bands, and finally (by substantially increasing the total time on target), in all three bands. However, a close examination of the measured polarization values obtained during the three measurements does not reveal any clear temporal  variation between them within the measurement errors. It appears that all three bands report polarization values of the same sign, but the {\it V'}-band polarization value is higher by about 50\,\% than the values found in the other two bands. The highest Stokes $V/I$ value we  found is just below 0.5\,\%.

\subsubsection{WD\,0745$+$115}

Like several of the targets in this section, WD\,0745$+$115 is a very recent addition to the list of relatively nearby WDs, although it lies outside the 40 pc volume. It was identified as a nearby WD by \citet{Jimenez-Esteban2018}, an identification confirmed by \citet{Genetal19}. A spectrum was obtained by \citet{Kiletal20}, who identified the WD as a DC star. The basic parameters were derived by \citet{Genetal19, Genetal21}, who reported that it has a relatively high mass of $0.93\,M_\odot$ and an effective temperature $T_{\rm eff}$ close to 10'000\,K. For this WD, the absence of Balmer lines is a very clear indication of a He-rich atmosphere. 

Strong circular polarization is present in all observed bands. The polarization approaches +2\% in the {\it V'} band, is close to +1\% in the {\it R'} band, and reverses sign and exceeds $-1$\% in the {\it B'} band. There is no significant indication of variability in the four measurements taken during about one year.

\subsubsection{WD\,0756$+$437}

WD\,0756$+$437 was observed as an extremely interesting target of opportunity that was easily included in our general program. The star was originally identified as a DF-C WD with a possible $\lambda$4670\,\AA\ feature \citep{Greetal77,Greenstein1979}. It was repeatedly observed photometrically and spectroscopically since then, especially after the polarimetric discovery of a large magnetic field (see below). \citet{Putney95} identified the object as a DA WD with a H-rich atmosphere on the basis of the comparison of features in the polarization spectrum with the theoretical spectrum of H. The star is within the 40\,pc volume around the Sun \citep{McCetal20,OBretal24}.  

An extremely high level of circular polarization that rises to the huge value of $V/I \sim 8$\,\% at around 5500\,\AA\ was discovered by \citet{Putney95}, who published a single (nearly featureless) flux spectrum and an excellent circular polarization spectrum. The observed  polarization level indicates that  this WD possesses a very strong field. Our broad-band Stokes $V/I$ measurements are fully consistent with Putney's data. 

Putney compared narrow polarization features with predicted wavelength positions of various components of H$\alpha$ in fields of hundreds of MG \citep{Wunetal85} and concluded that the mean (\bs) field must be about 200--220\,MG in strength.  This field strength was also supported by \citet{Schetal03} on the basis of SDSS intensity spectra.  Comparison of the spectral energy distribution with models \citep{Limetal15,Genetal21} suggests that $\teff = 7215$\,K, $M = 1.04 M_\odot$, and that the cooling age is about 4.45\,Gyr. This DAH WD is thus probably (one of?) the oldest known, strongly magnetized products of a WD-WD merger \citep{Garetal12,BagLan22}.

We took five polarization data sets during two years. From the fact that the first two measurements, taken about 3\,hr apart, also provide the largest and smallest polarization values in our measurement set,  we deduce that the rotation period of this star could be of the order of six hours. Photometry by \citet{Brietal13} showed very large amplitude sinusoidal photometric variations (semi-amplitude of $\pm 4$\,\%) with a unique period of 6.68\,hr. Because the polarimetric variations are due to and locked to the stellar rotation, the agreement of the polarimetric period constraints with the photometric period confirms that the photometric period is the stellar rotation period.

%%%It is rapidly variable ($P~$ hours).

\subsection{Other targets with marginal detections}
In addition to these reasonably certain detections, the observations of nine other DC WDs provided a single detection at or slightly above the $3\sigma$ level in one band. These WDs can be regarded as candidates that need further measurements for a confirmation (for these targets, only the significant detections, but not the names, are shown in boldface in Table\,\ref{Tab_WD_Log}). They are 
WD\,0052$+$595; 
WD\,0102$+$210; 
WD\,0156$+$155;
WD\,0357$+$513;
WD\,0407$+$197;
WD\,0426$+$588; 
WD\,1554$-$079;
WD\,1731$+$290. 
Because of previous experience with unsuccessfully trying to confirm magnetic fields based on a single $3 \sigma$ polarization detection (see Sect.~\ref{Sect_Not_Confirmed} below), we suspect that the majority of these detections will not be confirmed, except perhaps for the case of WD\,0052$+$595.

\subsection{Targets with unconfirmed detections}\label{Sect_Not_Confirmed}
In our previous observing runs, polarization was detected once in a single band at a level slightly higher than $3\sigma$ in observations of WD1434$+$437 and WD\,1533$+$469 \citep{Beretal23}. We obtained one more observation of each of these two stars that failed to confirm the marginal detections reported earlier. We dropped these stars from our list of DCH candidates.

%%%%%%%%%%%%%%%%%%%%%
\section{Discussion and conclusions}
%%%%%%%%%%%%%%%%%%%%%

%%%%%%%%%%%%%%%%%%%%%
%%%\input{Tab_DCH_Summary}
%%%%%%%%%%%%%%%%%%%%%
The circular polarization levels detected with the DIPoL-UF at NOT range from about 1\,\%, found in two DC WDs to a number of detections of polarization at levels between 0.2\,\% and 0.6\,\%, with measurement uncertainties ranging between about 0.02\,\% to 0.1\,\%. It is clear that an extremely important feature of the DIPoL-UF polarimeter is its stable and precisely measured zeropoint \citep{Beretal22}. However, our observing program reaches the limit of what can be achieved with a medium-sized telescope such as the 2.5 m NOT. Further substantial progress will require the use of larger telescopes.  

Figure~\ref{Fig_ETC} shows the relation of the magnitude and the uncertainty of our measurement errors of all NOT+DIPoL-UF measurements of WDs as a function of \textit{Gaia} $G$ magnitude, normalized to 1\,h exposure time. This figure represents a substantial update compared to Fig.~4 of \citet{Beretal22}. It may be used to make quantitative predictions regarding the level of polarimetric precision that may be achieved for the BBCP measurements as a function of exposure time and stellar magnitude. It can also be used to evaluate observational requirements for a given WD with DIPol-UF employed on a larger telescope.

Our search for broad-band circular polarization, and therefore, for magnetic fields, in DC WDs continues to deliver detections of previously unknown fields in these faint and featureless objects, with a detection rate of 13 new MWDs so far out of 101 observed new targets. 
The diverse outcome obtained in our different runs is remarkable: \citet{Beretal22} reported the detection of 2 new MWDs out of 3 newly observed DC WDs, \citet{Beretal23} reported 5 out of 23, and we detected and reported 6 new MWDs out of 75. {Overall, the detection of a magnetic field in 13 out of 105 observed WDs is fully consistent with the frequency of the occurrence of magnetic fields in the DC WDs in the local 20\,pc volume, $13.3 \pm 6.2$\%, that was measured by \citet{BagLan21}. This value is certainly an underestimate of the actual frequency of magnetic fields in old WDs because stars with magnetic fields weaker than a few MG do not produce a signal of circular polarization that can be detected with our instruments. On the other hand,} only 2 of the newly discovered magnetic DC stars show circular polarization levels in excess of 1\,\%. We found no new magnetic DC WDs with extremely high polarization levels in our survey (several percent polarization), similar to those found in other types of magnetic WDs, such as WD\,1900+705 (=Grw+70$^\circ$\,8247, 3\%), WD\,0756+437 \citep[=EGGR\,428, $\sim 8$\,\%][]{Kemetal70}, CL\,Oct \citep[=EUVE J0317--85.5, $\sim 11$\%][]{Baretal95}, and WD\,1031+234 \citep[=PG\,1031+234, $\sim 12$\%][]{WesSch84,Latetal86,Schetal86}. {Our new data, combined with spectropolarimetric surveys of WDs of various cooling ages, will eventually help us to understand the origin and evolution of magnetic fields in degenerate stars.}

We note that roughly half of the DCs observed in this survey belong to the category of younger DCs that have He atmospheres, $\teff \ga 6000$\,K, and cooling ages younger than about 4\,Gyr. These stars were presumably DB WDs earlier in their cooling. The other, older, DCs, also half of the sample, with lower \teff\ values and ages older 4 or 5\,Gyr, may have H-rich or He-rich atmospheres, and these stars sample the production of magnetic WDs during the first few billion years of the evolution of the Milky Way galaxy. 

We cannot assign definite magnetic field strengths to the observed polarization signals because the theory of continuum circular polarization in these very cool WDs has not been studied in detail.  However, using a rough estimate (derived from somewhat hotter magnetic WDs) that polarization $V/I \sim 1$\,\% is produced by a mean line-of-sight field component \bz\ of about 15\,MG \citep{BagLan20}, most of the fields detected in our survey probably have overall typical field strengths \bs\ of a few MG to tens of MG, rising to perhaps as high as 100\,MG in the most highly polarized WDs. 

The magnetic fields of DC WDs has been an almost completely neglected topic since the 1970s, when searches with broad-band polarimetry went out of fashion. We showed that this method of searching for fields in these generally faint WDs is still an extremely powerful tool that has revealed MG-level fields in a significant fraction of the DCs. These newly discovered magnetic DC WDs can now be studied with spectropolarimetry using larger telescopes, which will provide additional constraints for developing a suitable model for their enigmatic circularly polarized continuum spectra.

\begin{acknowledgements}
Based on observations made with the Nordic Optical Telescope, owned in collaboration by the University of Turku and Aarhus University, and operated jointly by Aarhus University, the University of Turku and the University of Oslo, representing Denmark, Finland and Norway, the University of Iceland and Stockholm University at the Observatorio del Roque de los Muchachos, La Palma, Spain, of the Instituto de Astrofisica de Canarias. DIPol-UF is a joint effort between University of Turku (Finland) and Leibniz Institute for Solar Physics (Germany). We  acknowledge  support from the Magnus Ehrnrooth foundation and the ERC Advanced Grant HotMol  ERC-2011-AdG-291659.
JDL acknowledges the financial support of the Natural Sciences and Engineering Research Council of Canada (NSERC), funding reference number 6377-2016.
\end{acknowledgements}

\bibliographystyle{aa}
\bibliography{sbabib}
%\begin{appendix}

%%%%%%%%%%%%%%%%%%%%%
%Added by TeX Support
\begin{onecolumn}
\begin{center}
%                 12345678
\begin{longtable}{llrrrccl}
\caption{\label{Table_Program} Basic parameters of stars observed.}\\
\hline\hline
\multicolumn{2}{c}{STAR}               &  $G$ &\multicolumn{1}{c}{$d$}& \teff &    $M$      & Age  & Atmosphere %%%and ref. 
\\
               &                       &      &(pc)                   & (K)   & ($M_\odot$) &(Gyr) &                     \\
%-------------
\hline
\endfirsthead
\caption{continued.}\\
\hline\hline
\multicolumn{2}{c}{STAR}               &  $G$ &\multicolumn{1}{c}{$d$}& \teff &    $M$      & Age  & Atmosphere %%%and ref. 
\\
               &                       &      &(pc)                   & (K)   & ($M_\odot$) &(Gyr) & \\
\hline
\endhead
\hline
\endfoot
{\bf WD\,0004$+$122  }& LP 464-57                    & 16.3 & 17.5 &   5223 & 0.76 & 7.614 & H      \\%  2766234439302571904 
{\bf WD\,0004$+$337 *}& LP 240-29                    & 17.2 & 34.9 &   5792 & 0.76 & 4.923 & He     \\%  2875903332533220992
{\rm WD\,0019$+$264  }& LP 349-1                     & 17.5 & 32.9 &   4991 & 0.58 & 5.974 & H      \\%  2855790657816672000
{\bf WD\,0019$+$423 *}& EGGR 459                     & 16.4 & 34.5 &   5625 & 0.39 & 2.055 & He     \\%  385105360675267840
{\bf WD\,0021$+$683 *}& G 242-54B                    & 17.5 & 36.0 &   5644 & 0.73 & 5.578 & He     \\%  529594417061837824
{\rm WD\,0034$-$285  }& 2MASS J00371373-2814503      & 17.6 & 37.7 &   5472 & 0.72 & 5.557 & H      \\%  2319052851148048256 
{\rm WD\,0052$+$595  }& LSPM J0055+5948              & 16.8 & 22.8 &   4645 & 0.48 & 5.694 & H      \\%  426122397136335872  
{\rm WD\,0102$+$210  }& EGGR 463                     & 17.8 & 32.5 &   4721 & 0.62 & 8.125 & H      \\%  2790494850219788160 
{\rm WD\,0156$+$155  }& PG 0156+156                  & 15.8 & 38.2 &   9507 & 0.76 & 1.132 & He     \\%  78629306318257536
{\rm WD\,0158$+$119  }& GD 21                        & 16.1 & 31.5 &   7224 & 0.68 & 1.926 & He     \\%  2574620550768898432 
{\rm WD\,0200$-$127  }& GD 1072                      & 14.5 & 24.1 &  10261 & 0.76 & 0.916 & He     \\%  5149836834977282048 
{\rm WD\,0222$+$647  }& LP 53-7                      & 17.9 & 31.0 &   4189 & 0.47 & 7.07  & H      \\%  515289392829738624  
{\rm WD\,0228$+$079  }& LP 530-21                    & 17.2 & 38.9 &   5850 & 0.60 & 3.322 & He     \\%  19693180966870656
{\rm WD\,0230$+$212  }& LP 410-52                    & 17.2 & 28.5 &   4730 & 0.50 & 5.60  & H      \\%  88996326578456192   
{\rm WD\,0306$+$663  }& LSR J0310+6634               & 17.9 & 38.4 &   4681 & 0.47 & 5.269 & H      \\%  492442090962517376 
{\rm WD\,0313$+$393  }& GD 44                        & 15.7 & 37.9 &   9162 & 0.72 & 1.123 & He     \\%  235842052999634944
{\rm WD\,0315$+$423  }& UCAC4 663-016444             & 14.8 & 30.6 &  11260 & 0.76 & 0.719 & He (DBA)\\%  239721228805415296  
{\rm WD\,0317$+$233  }& LP 355-59                    & 17.6 & 39.8 &   4690 & 0.40 & 3.941 & H      \\%  62884540326096896
{\rm WD\,0342$-$039  }& 2MASS J03450153-0348492      & 17.8 & 31.0 &   4362 & 0.49 & 6.907 & H      \\%  3249479592234269056
{\rm WD\,0344$+$014  }& LP 593-56                    & 16.3 & 20.2 &   4959 & 0.51 & 4.530 & H      \\%  3270079526697712768 
{\rm WD\,0357$+$513  }& LSR J0401+5131               & 17.1 & 25.1 &   5018 & 0.69 & 6.975 & He     \\%  250862824946594816  
{\rm WD\,0359$+$153  }& 2MASS J04024236+1527419      & 16.6 & 36.9 &   7145 & 0.75 & 2.653 & He     \\%  39387328302768640 
{\rm WD\,0407$+$197  }& LP 414-106                   & 17.5 & 29.1 &   4965 & 0.69 & 8.200 & H      \\%  48829075167625472   
{\bf WD\,0419$+$576 *}& LP 84-16                     & 16.6 & 37.4 &   7439 & 0.77 & 2.473 & He     \\%  470639806179201792
{\rm WD\,0426$+$588  }& EGGR 180                     & 12.3 &  5.5 &   7294 & 0.72 & 2.187 & He     \\%  470826482637310848   
{\rm WD\,0431$+$308  }& LP 302-33                    & 17.8 & 34.1 &   4764 & 0.60 & 7.658 & H      \\%  159277625222514048
{\rm WD\,0535$+$438  }& LP 203-56                    & 17.3 & 39.3 &   5961 & 0.67 & 3.872 & He     \\%  197048460976343168
{\rm WD\,0541$+$260  }& LSR J0544+2603               & 16.9 & 36.0 &   4778 & 0.26 & 2.192 & H      \\%  3429296884940000000
{\rm WD\,0543$+$112  }& LSPM J0546+1115              & 17.7 & 36.5 &   4870 & 0.53 & 5.483 & H      \\%  3337021260634868224 
{\bf WD\,0548$-$001  }& EGGR 248                     & 14.4 & 11.2 &   6053 & 0.65 & 3.35  & He (DQ)\\% 
{\rm WD\,0549$-$261  }&  2MASS J05511869-2609125     & 17.2 & 39.6 &   4752 & 0.27 & 2.373 & H      \\%  2914000658819492864
{\rm WD\,0638$-$044  }& PM J06411-0432               & 17.0 & 20.4 &   4062 & 0.45 & 7.092 & H      \\%  3103811515783541504
{\bf WD\,0654$+$059 *}& 2MASS J06572938+0550479      & 17.3 & 39.0 &   6010 & 0.69 & 3.961 & He     \\%  3129655296079487872
{\rm WD\,0714$+$458  }& GD 84                        & 15.2 & 35.5 &  10314 & 0.67 & 0.751 & He     \\%  974375354722176768
{\rm WD\,0726$+$536  }& LP 123-5                     & 17.9 & 36.2 &   4589 & 0.51 & 6.573 & H      \\%  983979721933760768 
{\bf WD\,0745$+$115 *}& GALEX J074842.4+112502       & 16.5 & 46.7 &   9831 & 0.93 & 1.708 & He     \\%  3151443115414582016
{\rm WD\,0748$+$614  }& GD 453                       & 16.0 & 35.0 &   8270 & 0.74 & 1.536 & He     \\%  1084764020047940096
{\rm WD\,0749$+$426  }& LP 207-50                    & 17.2 & 27.4 &   4585 & 0.47 & 5.576 & H      \\%  925532265076649472 
{\rm WD\,0751$+$578  }& G 193-78                     & 15.0 & 29.4 &  10011 & 0.76 & 0.980 & He     \\%  1081813343155426560
{\bf WD\,0756$+$437  }& EGGR 428                     & 16.1 & 22.0 &   7215 & 1.04 & 4.450 & H      \\%  923023248261420672 
{\rm WD\,0759$-$283  }& Gaia DR3 5597759970724418688 & 16.8 & 35.0 &   5682 & 0.48 & 2.502 & He (DQp)\\% 5597759970724418688
{\rm WD\,0810$+$489  }& G 111-64                     & 14.9 & 17.1 &   6709 & 0.65 & 2.17  & He     \\%  931573222477949696  
{\rm WD\,0847$-$017  }& LP 606-12                    & 17.1 & 27.9 &   4884 & 0.54 & 5.727 & H      \\%
{\rm WD\,0848$+$166  }& LP 426-17                    & 16.6 & 30.9 &   5907 & 0.59 & 3.024 & He     \\%  611645983387812992  
{\rm WD\,0853$-$083  }& PM J08555-0833               & 17.6 & 33.1 &   4579 & 0.46 & 5.544 & H      \\%  5755957119598921728
{\rm WD\,0900$+$734  }&  G 252-27                    & 16.8 & 25.7 &   5065 & 0.53 & 5.09  & He     \\%  1123700235048742016
{\rm WD\,0945$+$206  }& LP 428-41                    & 17.8 & 37.2 &   4909 & 0.58 & 6.401 & H      \\%  639665392247496576
{\rm WD\,0947$+$153  }& LP 428-53                    & 17.4 & 39.2 &   5320 & 0.51 & 3.950 & He     \\%  616396182856185728
{\rm WD\,0958$+$436  }& SDSS J100204.08+432645.6     & 17.8 & 34.9 &   4699 & 0.52 & 6.251 & H      \\%  808030648576069760
{\bf WD\,1008$+$290  }& LP 315-42                    & 16.5 & 14.7 &   3962 & 0.41 & 5.991 & He (DQp)\\%  740483560857296768 
{\rm WD\,1012$+$083  }& LP 549-32                    & 17.5 & 29.5 &   4558 & 0.50 & 6.395 & H      \\%  3875651975353757440 
{\bf WD\,1036$-$204  }& EGGR 535                     & 15.8 & 14.1 &   5754 & 0.94 & 6.185 & He (DQp)\\% 3553682127126319360  
{\rm WD\,1037$+$103  }& PM J10403+1004               & 17.1 & 36.9 &   5983 & 0.65 & 3.614 & He      \\%  3870354528331257984
{\rm WD\,1049$+$411  }& LP 213-79                    & 16.7 & 33.9 &   6237 & 0.64 & 2.815 & He     \\%  776762672481353984
{\rm WD\,1055$-$072  }& LAWD\,34                     & 14.2 & 12.3 &   7388 & 0.81 & 2.822 & He     \\%  3763445409285757824  
{\rm WD\,1122$+$214  }& LP 374-46                    & 16.9 & 38.1 &   6194 & 0.60 & 2.410 & He     \\%  3978862277154958592 
{\rm WD\,1143$+$633  }&  EGGR 353                    & 16.2 & 24.0 &   5484 & 0.53 & 2.589 & H (DA) \\%  863131372427958912
{\rm WD\,1153$+$135  }& LP 493-78                    & 17.4 & 35.6 &   4769 & 0.42 & 3.740 & H      \\%  3920187251456610816
{\rm WD\,1155$+$003  }& LSPM J1158+0004              & 17.9 & 34.6 &   4415 & 0.46 & 6.182 & H      \\%  3795052348495488896   
{\rm WD\,1223$+$188  }& LP 435-109                   & 16.2 & 35.9 &   7972 & 0.80 & 2.031 & H (DAHe) \\%  3947104533054775168    
{\rm WD\,1239$-$133  }& EC 12393-1318                & 15.7 & 35.9 &   8643 & 0.68 & 1.196 & He     \\%  3528871819044810368
{\rm WD\,1245$-$102  }& 2MASS J12482817-1028576      & 16.8 & 39.7 &   7398 & 0.82 & 2.854 & He     \\%  3530520910392199680 
{\rm WD\,1310$+$027  }& LP 557-24                    & 17.9 & 30.9 &   4165 & 0.47 & 7.154 & H      \\%  3691685882071367936
{\bf WD\,1315$+$222  }& LP 378-956                   & 16.7 & 31.8 &   6334 & 0.72 & 3.541 & He     \\%  3943650619138622848 
{\rm WD\,1343$+$422  }& LP 219-63                    & 17.0 & 37.5 &   4639 & 0.24 & 2.270 & H      \\%  1500607765872799616
{\bf WD\,1346$+$121  }& LP 498-66                    & 17.8 & 28.3 &   4435 & 0.63 & 9.231 & H      \\%  3728074738695246336
{\rm WD\,1419$+$426  }& PB 2202                      & 17.3 & 45.8 &   8901 & 1.12 & 3.236 & H (DA) \\%  1492278518615941504 
{\rm WD\,1434$+$437  }&  LP 221-217                  & 17.2 & 27.2 &   4634 & 0.49 & 5.886 & H      \\%  1493367245581725184
{\rm WD\,1533$+$469  }&  LP 176-60                   & 17.8 & 30.8 &   4309 & 0.47 & 6.678 & H      \\%  1401010605309570816
{\rm WD\,1547$+$572  }&  LP 99-578                   & 17.3 & 37.9 &   5770 & 0.67 & 4.567 & He     \\%  1598771000065137536
{\rm WD\,1554$-$079  }& NAME SSS J1556-0806          & 18.3 & 32.7 &   7155 & 1.29 & 2.712 & He     \\%  4348098485293072128 
{\bf WD\,1556$+$044  }& PM J15589+0417               & 16.0 & 22.5 &   6867 & 0.91 & 4.218 & H      \\%  4425632987265111680 
{\rm WD\,1717$+$104  }& 2MASS J17200675+1022278      & 17.6 & 37.6 &   4973 & 0.51 & 5.122 & He     \\%  4491748511228743808
{\rm WD\,1731$+$290  }& LSPM J1733+2903              & 17.0 & 39.6 &   6458 & 0.69 & 2.961 & He     \\%  4598775557191664384
{\rm WD\,1741$+$145  }& LSPM J1743+1434S             & 14.9 & 34.3 &  10572 & 0.60 & 0.598 & He     \\%  4500646618315862144
{\rm WD\,1745$+$290  }&  LSPM J1747+2859             & 17.5 & 36.6 &   4832 & 0.45 & 3.974 & H      \\% 4596322473734130304
{\bf WD\,2049$-$223  }& LP 872-48                    & 14.9 & 20.3 &   8145 & 0.78 & 1.904 & He     \\% 6808651507904773888 
{\bf WD\,2049$-$252  }& UCAC4 325-215293             & 16.0 & 18.0 &   4908 & 0.49 & 4.540 & H      \\% 6805792571514600960  
{\bf WD\,2211$+$372  }&  LP 287-35                   & 16.8 & 29.2 &   6424 & 0.88 & 4.379 & He     \\% 1955134710179436672 
{\rm WD\,2228$+$226  }&  2MASS J22305914+2254543     & 17.0 & 31.0 &   5774 & 0.70 & 4.841 & He     \\%  1875301369907249024 
{\rm WD\,2303$+$391  }& LSPM J2305+3922              & 17.9 & 35.9 &   7021 & 1.17 & 3.806 & He     \\%  1929838143078434432 
{\rm WD\,2345$+$027  }& LP 643-65                    & 16.6 & 25.2 &   4994 & 0.47 & 3.520 & H      \\%  2742789930821144320
{\rm WD\,2359$+$636 }&  LSR J0002+6357               & 17.0 & 26.3 &   4879 & 0.54 & 5.811 & H      \\%  431635455820288128  
\hline
\end{longtable}
\end{center}
\noindent
Detections are marked in boldface. Newly discovered magnetic WDs are marked with a *.
All stars for which no spectral type is explicitly given in the last column have a featureless spectrum (spectral class DC).\\
%%%Key to references: 
%%%1: \citet{Bloetal19};
%%%2: \citet{McCetal20};
%%%2: \citet{Genetal21};
%%%4: \citet{Treetal20};
%%%3: \citet{Beretal22}.
Chemical composition of the stellar atmosphere was taken from \citet{OBretal24}, temperature and mass from \citet{Genetal21}, and
ages were interpolated using the tables of \citet{Bedetal20}.
\end{onecolumn}

%%%%%%%%%%%%%%%%%%%%%

%%%%%%%%%%%%%%%%%%%%%
%Added by TeX Support
%%%%\longtab{

%                 123456        78        90        1
\begin{center}
\begin{onecolumn}
\begin{longtable}{lcccrr@{$\pm$}lr@{$\pm$}lr@{$\pm$}l}
\caption{\label{Tab_WD_Log} Observing log of WDs. Detections are marked in boldface. Newly discovered magnetic stars are marked with a *.}\\
\hline\hline
\multicolumn{1}{c}{STAR} &    DATE  &  UT &JD --  & Exp. & \multicolumn{6}{c}{$V/I$ (\%)} \\
                         &yyyy-mm-dd&hh:mm&2400000&\multicolumn{1}{c}{(s)}  &
                         \multicolumn{2}{c}{$B'$}&\multicolumn{2}{c}{$V'$} &\multicolumn{2}{c}{$R'$} \\
%-------------
\hline
\endfirsthead
\caption{continued.}\\
\hline\hline
\multicolumn{1}{c}{STAR} &    DATE  &  UT &JD --  & Exp. & \multicolumn{6}{c}{$V/I$ (\%)} \\
                         &yyyy-mm-dd&hh:mm&2400000&\multicolumn{1}{c}{(s)}  &
                         \multicolumn{2}{c}{$B'$}&\multicolumn{2}{c}{$V'$} &\multicolumn{2}{c}{$R'$} \\
%-------------
\hline
\endhead
\hline
\endfoot
{\bf WD\,0004$+$122  }& 2022-07-01 & 04:42 &  59761.696 & 3120 &${\bf  0.813 }$&{\bf 0.060 }&${\bf  0.761 }$&{\bf 0.052 }&${\bf  1.268 }$&{\bf 0.037 }\\
                      & 2022-07-06 & 02:36 &  59766.608 & 3360 &${\bf  0.838 }$&{\bf 0.056 }&${\bf  0.676 }$&{\bf 0.042 }&${\bf  1.181 }$&{\bf 0.038 }\\
                      & 2022-11-20 & 23:27 &  59904.477 & 3200 &${\bf  0.982 }$&{\bf 0.049 }&${\bf  0.731 }$&{\bf 0.063 }&${\bf  1.201 }$&{\bf 0.047 }\\
                      & 2023-11-16 & 22:16 &  60265.428 & 3120 &${\bf  0.814 }$&{\bf 0.066 }&${\bf  0.794 }$&{\bf 0.051 }&${\bf  1.187 }$&{\bf 0.045 }\\ [1mm]
{\bf WD\,0004$+$337 *}& 2023-11-17 & 20:25 &  60266.351 & 4800 &${\bf  0.644 }$&{\bf 0.070 }&${\bf -0.154 }$&{\bf 0.050 }&${\bf  0.451 }$&{\bf 0.056 }\\ [1mm]  
{\rm WD\,0019$+$264  }& 2022-11-21 & 20:43 &  59905.364 & 4400 &${\rm  0.154 }$&{\rm 0.086 }&${\rm -0.053 }$&{\rm 0.062 }&${\rm  0.027 }$&{\rm 0.069 }\\ [1mm]  
{\bf WD\,0019$+$423 *}& 2023-11-17 & 22:16 &  60266.405 & 3120 &${\bf  0.814 }$&{\bf 0.066 }&${\rm  0.000 }$&{\rm 0.047 }&${\rm  0.026 }$&{\rm 0.049 }\\ [1mm]  
{\bf WD\,0021$+$683 *}& 2023-11-19 & 20:33 &  60268.356 & 5600 &${\rm  0.222 }$&{\rm 0.118 }&${\bf  0.292 }$&{\bf 0.083 }&${\bf -0.435 }$&{\bf 0.075 }\\ [1mm]  
{\rm WD\,0034$-$285  }& 2023-11-18 & 20:41 &  60267.362 & 5600 &${\rm  0.283 }$&{\rm 0.126 }&${\rm  0.015 }$&{\rm 0.101 }&${\bf -0.364 }$&{\bf 0.116 }\\ [1mm]  
{\rm WD\,0052$+$595  }& 2022-11-19 & 00:15 &  59902.512 & 4160 &${\rm  0.039 }$&{\rm 0.077 }&${\bf -0.298 }$&{\bf 0.099 }&${\bf -0.203 }$&{\bf 0.063 }\\
                      & 2022-11-21 & 22:09 &  59905.423 & 3840 &${\rm -0.045 }$&{\rm 0.086 }&${\rm -0.028 }$&{\rm 0.067 }&${\rm  0.080 }$&{\rm 0.041 }\\
                      & 2023-11-16 & 23:46 &  60265.490 & 5440 &${\rm  0.067 }$&{\rm 0.124 }&${\rm -0.014 }$&{\rm 0.103 }&${\bf  0.291 }$&{\bf 0.086 }\\ [1mm]
{\rm WD\,0102$+$210  }& 2022-11-21 & 00:58 &  59904.540 & 6240 &${\rm -0.002 }$&{\rm 0.174 }&${\rm  0.247 }$&{\rm 0.169 }&${\bf -0.234 }$&{\bf 0.068 }\\
                      & 2023-11-18 & 22:35 &  60267.441 & 6000 &${\rm -0.134 }$&{\rm 0.091 }&${\rm  0.175 }$&{\rm 0.092 }&${\rm  0.018 }$&{\rm 0.068 }\\ [1mm] 
{\rm WD\,0156$+$155  }& 2024-02-05 & 21:22 &  60346.390 & 2400 &${\rm -0.044 }$&{\rm 0.039 }&${\rm -0.030 }$&{\rm 0.070 }&${\bf  0.193 }$&{\bf 0.046 }\\ [1mm]
{\rm WD\,0158$+$119  }& 2022-11-19 & 01:29 &  59902.563 & 3120 &${\rm  0.073 }$&{\rm 0.044 }&${\rm  0.065 }$&{\rm 0.042 }&${\rm -0.106 }$&{\rm 0.042 }\\ [1mm]
{\rm WD\,0200$-$127  }& 2024-02-04 & 20:25 &  60345.351 & 1200 &${\rm  0.044 }$&{\rm 0.016 }&${\rm -0.015 }$&{\rm 0.030 }&${\rm  0.003 }$&{\rm 0.026 }\\ [1mm]
{\rm WD\,0222$+$647  }& 2022-11-21 & 23:53 &  59905.495 & 6240 &${\rm  0.089 }$&{\rm 0.159 }&${\rm  0.199 }$&{\rm 0.098 }&${\rm -0.104 }$&{\rm 0.074 }\\ [1mm]
{\rm WD\,0228$+$079  }& 2023-11-17 & 23:03 &  60266.461 & 4800 &${\rm -0.104 }$&{\rm 0.049 }&${\rm  0.002 }$&{\rm 0.053 }&${\rm -0.029 }$&{\rm 0.047 }\\ [1mm]
{\rm WD\,0230$+$212  }& 2022-11-20 & 02:20 &  59903.597 & 5200 &${\bf -0.342 }$&{\bf 0.112 }&${\rm -0.119 }$&{\rm 0.068 }&${\rm  0.024 }$&{\rm 0.053 }\\ [1mm] 
{\rm WD\,0306$+$663  }& 2023-11-19 & 01:59 &  60267.582 & 6400 &${\rm  0.095 }$&{\rm 0.109 }&${\rm  0.089 }$&{\rm 0.123 }&${\rm  0.044 }$&{\rm 0.076 }\\ [1mm] 
{\rm WD\,0313$+$393  }& 2023-11-17 & 01:20 &  60265.556 & 3360 &${\rm -0.053 }$&{\rm 0.030 }&${\rm -0.000 }$&{\rm 0.043 }&${\rm  0.017 }$&{\rm 0.042 }\\ [1mm]
{\rm WD\,0315$+$423  }& 2022-11-20 & 03:28 &  59903.645 & 1760 &${\rm -0.024 }$&{\rm 0.028 }&${\rm -0.031 }$&{\rm 0.021 }&${\rm  0.032 }$&{\rm 0.035 }\\ [1mm]
{\rm WD\,0317$+$233  }& 2024-02-06 & 21:19 &  60347.388 & 6240 &${\rm -0.349 }$&{\rm 0.123 }&${\rm -0.120 }$&{\rm 0.102 }&${\rm -0.123 }$&{\rm 0.072 }\\ [1mm]
{\rm WD\,0342$-$039  }& 2024-02-07 & 21:19 &  60348.388 & 6240 &${\rm  0.224 }$&{\rm 0.168 }&${\rm  0.042 }$&{\rm 0.087 }&${\rm -0.010 }$&{\rm 0.075 }\\ [1mm]
{\rm WD\,0344$+$014  }& 2022-11-19 & 02:32 &  59902.606 & 3120 &${\rm  0.019 }$&{\rm 0.068 }&${\rm  0.072 }$&{\rm 0.057 }&${\rm -0.022 }$&{\rm 0.038 }\\ [1mm]
{\rm WD\,0357$+$513  }& 2022-11-19 & 03:54 &  59902.663 & 4800 &${\rm -0.114 }$&{\rm 0.078 }&${\rm -0.031 }$&{\rm 0.087 }&${\bf -0.161 }$&{\bf 0.050 }\\ [1mm]
{\rm WD\,0359$+$153  }& 2023-11-19 & 00:13 &  60267.509 & 3840 &${\rm  0.054 }$&{\rm 0.047 }&${\rm -0.049 }$&{\rm 0.041 }&${\rm -0.044 }$&{\rm 0.049 }\\ [1mm]
{\rm WD\,0407$+$197  }& 2022-11-20 & 04:45 &  59903.698 & 5280 &${\bf  0.546 }$&{\bf 0.133 }&${\rm  0.170 }$&{\rm 0.125 }&${\rm -0.015 }$&{\rm 0.083 }\\ 
                      & 2022-11-21 & 03:01 &  59904.626 & 5280 &${\rm  0.040 }$&{\rm 0.119 }&${\rm -0.227 }$&{\rm 0.096 }&${\rm -0.026 }$&{\rm 0.080 }\\ [1mm]
{\bf WD\,0419$+$576 *}& 2023-11-18 & 00:30 &  60266.521 & 3600 &${\bf  0.157 }$&{\bf 0.047 }&${\bf -0.154 }$&{\bf 0.050 }&${\bf -0.165 }  $& {\bf 0.033}\\ [1mm]  
{\rm WD\,0426$+$588  }& 2024-02-06 & 22:34 &  60347.440 & 1200 &${\bf -0.042 }$&{\bf 0.010 }&${\rm  0.007 }$&{\rm 0.012 }&${\rm  0.028 }$&{\rm 0.014 }\\ [1mm]
{\rm WD\,0431$+$308  }& 2023-11-20 & 02:14 &  60268.593 & 5600 &${\rm -0.423 }$&{\rm 0.162 }&${\rm  0.337 }$&{\rm 0.128 }&${\rm -0.317 }$&{\rm 0.109 }\\ [1mm]  
{\rm WD\,0535$+$438  }& 2023-03-24 & 21:37 &  60028.401 & 5040 &${\rm -0.107 }$&{\rm 0.075 }&${\rm -0.027 }$&{\rm 0.071 }&${\rm -0.056 }$&{\rm 0.065 }\\ [1mm]
{\rm WD\,0541$+$260  }& 2023-03-23 & 21:37 &  60027.401 & 3960 &${\rm  0.042 }$&{\rm 0.124 }&${\rm -0.102 }$&{\rm 0.085 }&${\rm -0.020 }$&{\rm 0.067 }\\ [1mm]
{\rm WD\,0543$+$112  }& 2023-03-26 & 21:41 &  60030.404 & 5600 &${\rm -0.000 }$&{\rm 0.223 }&${\rm -0.160 }$&{\rm 0.147 }&${\rm  0.081 }$&{\rm 0.124 }\\ [1mm]
{\bf WD\,0548$-$001  }& 2023-03-25 & 21:00 &  60029.375 & 1440 &${\bf -0.407 }$&{\bf 0.034 }&${\bf -0.807 }$&{\bf 0.037 }&${\bf -0.801 }$&{\bf 0.018 }\\ 
                      & 2023-11-17 & 02:31 &  60265.605 & 3120 &${\bf -0.394 }$&{\bf 0.024 }&${\bf -0.887 }$&{\bf 0.020 }&${\bf -0.821 }$&{\bf 0.029 }\\ [1mm]
{\rm WD\,0549$-$261  }& 2023-11-18 & 01:57 &  60266.581 & 4800 &${\rm -0.008 }$&{\rm 0.086 }&${\bf -0.191 }$&{\bf 0.050 }&${\rm  0.008 }$&{\rm 0.055 }\\ [1mm]  
{\rm WD\,0638$-$044  }& 2023-03-27 & 21:42 &  60031.404 & 4400 &${\rm  0.286 }$&{\rm 0.203 }&${\rm  0.066 }$&{\rm 0.101 }&${\rm -0.002 }$&{\rm 0.095 }\\ [1mm]
{\bf WD\,0654$+$059 *}& 2023-03-24 & 23:18 &  60028.471 & 4400 &${\rm -0.143 }$&{\rm 0.148 }&${\bf -0.463 }$&{\bf 0.089 }&${\rm -0.103 }$&{\rm 0.077 }\\ 
                      & 2023-03-25 & 22:12 &  60029.425 & 5200 &${\bf -0.279 }$&{\bf 0.093 }&${\bf -0.466 }$&{\bf 0.094 }&${\rm -0.219 }$&{\rm 0.078 }\\
                      & 2023-11-18 & 03:34 &  60266.649 & 4160 &${\bf -0.373 }$&{\bf 0.050 }&${\bf -0.491 }$&{\bf 0.055 }&${\bf -0.275 }$&{\bf 0.064 }\\ [1mm]
{\rm WD\,0714$+$458  }& 2024-02-04 & 21:10 &  60345.382 & 1680 &${\rm  0.022 }$&{\rm 0.031 }&${\rm  0.060 }$&{\rm 0.038 }&${\rm  0.039 }$&{\rm 0.034 }\\ [1mm]
{\rm WD\,0726$+$536  }& 2023-11-19 & 04:09 &  60267.6730& 5200 &${\rm  0.070 }$&{\rm 0.136 }&${\rm  0.048 }$&{\rm 0.114 }&${\rm  0.095 }$&{\rm 0.073 }\\ [1mm] 
{\bf WD\,0745$+$115 *}& 2023-03-23 & 23:10 &  60027.465 & 3120 &${\bf -1.089 }$&{\bf 0.053 }&${\bf  1.751 }$&{\bf 0.071 }&${\bf  0.874 }$&{\bf 0.065 }\\ 
                      & 2023-03-25 & 23:37 &  60029.484 & 3120 &${\bf -1.145 }$&{\bf 0.067 }&${\bf  1.680 }$&{\bf 0.064 }&${\bf  0.878 }$&{\bf 0.056 }\\
                      & 2023-11-17 & 03:48 &  60265.658 & 4080 &${\bf -1.124 }$&{\bf 0.063 }&${\bf  1.881 }$&{\bf 0.058 }&${\bf  0.906 }$&{\bf 0.052 }\\ [1mm] 
                      & 2024-02-04 & 22:18 &  60345.429 & 4800 &${\bf -1.165 }$&{\bf 0.046 }&${\bf  1.768 }$&{\bf 0.037 }&${\bf  0.895 }$&{\bf 0.043 }\\ [1mm] 
{\rm WD\,0748$+$614  }& 2023-03-26 & 23:03 &  60030.461 & 2880 &${\rm -0.007 }$&{\rm 0.062 }&${\rm -0.008 }$&{\rm 0.062 }&${\rm -0.018 }$&{\rm 0.040 }\\ [1mm]
{\rm WD\,0749$+$426  }& 2022-11-22 & 03:08 &  59905.631 & 5200 &${\rm  0.010 }$&{\rm 0.117 }&${\rm  0.153 }$&{\rm 0.079 }&${\rm -0.033 }$&{\rm 0.054 }\\ [1mm]
{\rm WD\,0751$+$578  }& 2022-11-20 & 05:50 &  59903.743 & 1600 &${\rm -0.004 }$&{\rm 0.030 }&${\rm -0.057 }$&{\rm 0.029 }&${\rm  0.023 }$&{\rm 0.033 }\\ [1mm]
{\bf WD\,0756$+$437  }& 2022-11-22 & 01:27 &  59905.561 & 3200 &${\bf  6.927 }$&{\bf 0.041 }&${\bf  8.558 }$&{\bf 0.064 }&${\bf  4.702 }$&{\bf 0.041 }\\
                      & 2022-11-22 & 04:31 &  59905.688 & 3200 &${\bf  4.458 }$&{\bf 0.029 }&${\bf  5.539 }$&{\bf 0.052 }&${\bf  3.798 }$&{\bf 0.044 }\\  
                      & 2024-02-03 & 21:59 &  60344.416 & 3680 &${\bf  5.041 }$&{\bf 0.117 }&${\bf  5.838 }$&{\bf 0.159 }&${\bf  4.452 }$&{\bf 0.159 }\\
                      & 2024-02-05 & 22:26 &  60346.435 & 3520 &${\bf  6.464 }$&{\bf 0.070 }&${\bf  8.187 }$&{\bf 0.092 }&${\bf  4.698 }$&{\bf 0.051 }\\
                      & 2024-02-06 & 23:24 &  60347.475 & 2880 &${\bf  5.108 }$&{\bf 0.034 }&${\bf  6.332 }$&{\bf 0.035 }&${\bf  4.406 }$&{\bf 0.040 }\\ [1mm]
{\rm WD\,0759$-$283  }& 2023-11-18 & 05:19 &  60266.721 & 4800 &${\rm -0.018 }$&{\rm 0.047 }&${\rm  0.206 }$&{\rm 0.081 }&${\bf  0.193 }$&{\bf 0.044 }\\ [1mm]  
{\rm WD\,0810$+$489  }& 2022-11-19 & 05:03 &  59902.711 & 1920 &${\rm -0.041 }$&{\rm 0.018 }&${\rm  0.001 }$&{\rm 0.023 }&${\rm -0.042 }$&{\rm 0.028 }\\ [1mm]
{\rm WD\,0847$-$017  }& 2024-02-05 & 05:46 &  60345.504 & 5600 &${\rm  0.169 }$&{\rm 0.070 }&${\rm -0.060 }$&{\rm 0.070 }&${\rm -0.054 }$&{\rm 0.048 }\\ [1mm]
{\rm WD\,0848$+$166  }& 2023-03-27 & 23:07 &  60031.463 & 4160 &${\rm -0.029 }$&{\rm 0.075 }&${\rm -0.078 }$&{\rm 0.047 }&${\rm -0.091 }$&{\rm 0.049 }\\ [1mm]
{\rm WD\,0853$-$083  }& 2024-02-05 & 02:04 &  60345.586 & 6400 &${\rm  0.051 }$&{\rm 0.096 }&${\rm -0.166 }$&{\rm 0.101 }&${\rm  0.172 }$&{\rm 0.062 }\\ [1mm]
{\rm WD\,0900$+$734  }& 2022-11-21 & 04:34 &  59904.690 & 4160 &${\rm  0.011 }$&{\rm 0.065 }&${\rm -0.073 }$&{\rm 0.056 }&${\rm -0.050 }$&{\rm 0.055 }\\ [1mm]
{\rm WD\,0945$+$206  }& 2024-02-06 & 00:22 &  60346.515 & 6720 &${\rm -0.004 }$&{\rm 0.095 }&${\rm -0.042 }$&{\rm 0.096 }&${\rm -0.182 }$&{\rm 0.063 }\\ [1mm]
{\rm WD\,0947$+$153  }& 2023-03-24 & 02:04 &  60027.586 & 5000 &${\rm -0.058 }$&{\rm 0.103 }&${\rm -0.150 }$&{\rm 0.090 }&${\rm  0.066 }$&{\rm 0.086 }\\ [1mm]
{\rm WD\,0958$+$206  }& 2024-02-06 & 02:37 &  60346.609 & 6240 &${\rm -0.093 }$&{\rm 0.092 }&${\rm -0.041 }$&{\rm 0.084 }&${\rm  0.168 }$&{\rm 0.057 }\\ [1mm]
{\bf WD\,1008$+$290  }& 2023-03-26 & 01:58 &  60029.582 & 3960 &${\bf -8.421 }$&{\bf 0.107 }&${\bf -7.037 }$&{\bf 0.109 }&${\bf  7.236 }$&{\bf 0.046 }\\
                      & 2023-03-28 & 01:06 &  60031.546 & 4320 &${\bf -8.585 }$&{\bf 0.112 }&${\bf -7.132 }$&{\bf 0.108 }&${\bf  7.308 }$&{\bf 0.044 }\\
                      & 2024-02-05 & 04:13 &  60345.676 & 4800 &${\bf -8.038 }$&{\bf 0.130 }&${\bf -7.242 }$&{\bf 0.114 }&${\bf  7.185 }$&{\bf 0.037 }\\
                      & 2024-02-08 & 05:15 &  60348.719 & 5120 &${\bf -8.367 }$&{\bf 0.118 }&${\bf -7.598 }$&{\bf 0.119 }&${\bf  7.151 }$&{\bf 0.041 }\\ [1mm]
{\rm WD\,1012$+$083  }& 2022-11-22 & 05:51 &  59905.744 & 4800 &${\rm  0.120 }$&{\rm 0.114 }&${\rm  0.040 }$&{\rm 0.081 }&${\rm -0.153 }$&{\rm 0.056 }\\ [1mm]
{\bf WD\,1036$-$204  }& 2022-11-21 & 05:42 &  59904.738 & 2640 &${\bf -4.450 }$&{\bf 0.060 }&${\bf -6.523 }$&{\bf 0.037 }&${\bf -9.196 }$&{\bf 0.032 }\\ 
                      & 2023-11-17 & 05:18 &  60265.721 & 4560 &${\bf -4.346 }$&{\bf 0.060 }&${\bf -6.441 }$&{\bf 0.055 }&${\bf -9.056 }$&{\bf 0.052 }\\
                      & 2024-02-08 & 03:45 &  60348.656 & 3120 &${\bf -4.598 }$&{\bf 0.047 }&${\bf -6.254 }$&{\bf 0.049 }&${\bf -9.219} $&{\bf 0.047 }\\ [1mm]
{\rm WD\,1037$+$103  }& 2023-03-27 & 01:24 &  60030.559 & 4480 &${\rm  0.075 }$&{\rm 0.067 }&${\rm  0.074 }$&{\rm 0.067 }&${\rm  0.021 }$&{\rm 0.044 }\\ [1mm]
{\rm WD\,1049$+$411  }& 2023-03-25 & 02:42 &  60028.613 & 3520 &${\rm -0.018 }$&{\rm 0.032 }&${\rm -0.014 }$&{\rm 0.061 }&${\rm  0.127 }$&{\rm 0.043 }\\ [1mm]
{\rm WD\,1055$-$072  }& 2022-11-19 & 05:45 &  59902.740 & 1920 &${\rm  0.022 }$&{\rm 0.021 }&${\rm  0.017 }$&{\rm 0.023 }&${\rm  0.023 }$&{\rm 0.023 }\\ [1mm]
{\rm WD\,1122$+$214  }& 2023-03-24 & 03:31 &  60027.647 & 3960 &${\rm  0.117 }$&{\rm 0.073 }&${\rm -0.200 }$&{\rm 0.068 }&${\rm  0.031 }$&{\rm 0.088 }\\ [1mm]
{\rm WD\,1143$+$633  }& 2023-11-19 & 04:09 &  60267.673 & 1680 &${\rm  0.070 }$&{\rm 0.136 }&${\rm -0.068 }$&{\rm 0.046 }&${\rm  0.032 }$&{\rm 0.027 }\\ [1mm]    
{\rm WD\,1153$+$135  }& 2023-03-25 & 04:06 &  60028.671 & 4800 &${\rm  0.263 }$&{\rm 0.131 }&${\rm  0.223 }$&{\rm 0.097 }&${\rm  0.083 }$&{\rm 0.085 }\\ [1mm]
{\rm WD\,1155$+$003  }& 2024-02-07 & 01:05 &  60347.545 & 7200&${\rm -0.297 }$&{\rm 0.103 }&${\rm  0.099 }$&{\rm 0.122 }&${\rm  0.031 }$&{\rm 0.062 }\\ [1mm]
{\rm WD\,1223$+$188  }& 2023-03-27 & 02:34 &  60030.607 & 2640 &${\rm  0.016 }$&{\rm 0.049 }&${\rm  0.009 }$&{\rm 0.050 }&${\rm -0.061 }$&{\rm 0.056 }\\ [1mm]
{\rm WD\,1239$-$133  }& 2024-02-05 & 06:24 &  60345.767 & 3120 &${\rm  0.052 }$&{\rm 0.029 }&${\rm -0.018 }$&{\rm 0.034 }&${\rm -0.037 }$&{\rm 0.032 }\\ [1mm]
{\rm WD\,1245$-$102  }& 2024-02-06 & 04:41 &  60346.695 & 4160 &${\rm  0.048 }$&{\rm 0.064 }&${\rm -0.053 }$&{\rm 0.056 }&${\rm  0.037 }$&{\rm 0.056 }\\ [1mm]
{\rm WD\,1310$+$027  }& 2022-07-03 & 22:41 &  59764.445 & 6000 &${\rm -0.300 }$&{\rm 0.133 }&${\rm -0.312 }$&{\rm 0.127 }&${\rm  0.081 }$&{\rm 0.084 }\\ [1mm]
{\bf WD\,1315$+$222  }& 2023-03-26 & 03:17 &  60029.637 & 3600 &${\bf  0.308 }$&{\bf 0.054 }&${\bf  0.224 }$&{\bf 0.078 }&${\bf  0.178 }$&{\bf 0.047 }\\ [1mm] 
{\rm WD\,1343$+$422  }& 2023-03-26 & 05:28 &  60029.728 & 4320 &${\rm -0.045 }$&{\rm 0.100 }&${\rm -0.013 }$&{\rm 0.090 }&${\rm  0.009 }$&{\rm 0.060 }\\ [1mm]
{\bf WD\,1346$+$121  }& 2023-03-27 & 04:10 &  60030.674 & 7100 &${\bf -0.415 }$&{\bf 0.137 }&${\bf -0.415 }$&{\bf 0.137 }&${\bf -1.187 }$&{\bf 0.070 }\\ [1mm]
{\rm WD\,1419$+$426  }& 2024-02-07 & 03:03 &  60347.627 & 4800 &${\rm -0.179 }$&{\rm 0.048 }&${\rm  0.095 }$&{\rm 0.055 }&${\rm -0.022 }$&{\rm 0.080 }\\ [1mm]
{\rm WD\,1434$+$437  }& 2024-02-06 & 06:13 &  60346.759 & 4160& ${\rm  0.150 }$&{\rm 0.069 }&${\rm -0.007 }$&{\rm 0.062 }&${\rm -0.042 }$&{\rm 0.050 }\\ [1mm]
{\rm WD\,1533$+$469  }& 2024-02-07 & 04:57 &  60347.706 & 6240 &${\rm  0.180 }$&{\rm 0.109 }&${\rm  0.174 }$&{\rm 0.115 }&${\rm  0.169 }$&{\rm 0.080 }\\ [1mm]
{\rm WD\,1547$+$572  }& 2023-03-28 & 02:38 &  60031.610 & 4800 &${\rm  0.034 }$&{\rm 0.078 }&${\rm -0.195 }$&{\rm 0.076 }&${\rm -0.096 }$&{\rm 0.077 }\\ [1mm]
{\rm WD\,1554$-$079  }& 2022-07-04 & 00:42 &  59764.529 & 6240 &${\bf -0.412 }$&{\bf 0.116 }&${\rm  0.019 }$&{\rm 0.108 }&${\rm  0.009 }$&{\rm 0.086 }\\ [1mm]
{\bf WD\,1556$+$044  }& 2022-06-28 & 23:39 &  59759.485 & 2880 &${\bf -0.262 }$&{\bf 0.038 }&${\bf  0.370 }$&{\bf 0.036 }&${\bf -0.160 }$&{\bf 0.042 }\\
                      & 2022-07-01 & 23:41 &  59762.487 & 3120 &${\bf -0.264 }$&{\bf 0.040 }&${\bf  0.401 }$&{\bf 0.038 }&${\bf -0.167 }$&{\bf 0.042 }\\
                      & 2023-03-26 & 04:17 &  60029.678 & 2400 &${\bf -0.294 }$&{\bf 0.044 }&${\bf  0.261 }$&{\bf 0.042 }&${\rm -0.119 }$&{\rm 0.057 }\\ [1mm] 
{\rm WD\,1717$+$104  }& 2023-03-24 & 05:19 &  60027.721 & 6720 &${\rm  0.001 }$&{\rm 0.139 }&${\rm  0.132 }$&{\rm 0.089 }&${\rm -0.011 }$&{\rm 0.086 }\\ [1mm]
{\rm WD\,1731$+$290  }& 2023-03-25 & 05:35 &  60028.733 & 3840 &${\bf -0.203 }$&{\bf 0.067 }&${\rm -0.074 }$&{\rm 0.068 }&${\rm -0.027 }$&{\rm 0.060 }\\ [1mm]
{\rm WD\,1741$+$145  }& 2024-02-07 & 06:22 &  60347.765 & 2240 &${\rm -0.037 }$&{\rm 0.026 }&${\rm -0.061 }$&{\rm 0.025 }&${\rm -0.030 }$&{\rm 0.027 }\\ [1mm]
{\rm WD\,1745$+$290  }& 2023-03-28 & 04:56 &  60031.706 & 5200 &${\rm  0.077 }$&{\rm 0.072 }&${\rm  0.202 }$&{\rm 0.097 }&${\rm  0.153 }$&{\rm 0.064 }\\ [1mm]
{\bf WD\,1900$+$705  }& 2023-11-16 & 20:00 &  60265.333 &  600 &${\bf 3.709  }$&{\bf 0.009 }&${\bf 3.499  }$&{\bf  0.016}&${\bf  3.784 }$&{\bf  0.016}\\     
{\bf WD\,2049$-$222  }& 2022-07-03 & 03:21 &  59763.639 & 1680 &${\bf  0.093 }$&{\bf 0.024 }&${\bf  0.122 }$&{\bf 0.023 }&${\rm  0.089 }$&{\rm 0.039 }\\ [1mm]
{\bf WD\,2049$-$253  }& 2022-07-01 & 03:03 &  59761.627 & 2640 &${\bf  0.358 }$&{\bf 0.056 }&${\bf  0.548 }$&{\bf 0.056 }&${\bf  0.584 }$&{\bf 0.035 }\\
                      & 2022-07-04 & 02:58 &  59764.624 & 3120 &${\bf  0.354 }$&{\bf 0.044 }&${\bf  0.551 }$&{\bf 0.045 }&${\bf  0.656 }$&{\bf 0.023 }\\ [1mm]
{\bf WD\,2211$+$372  }& 2022-11-18 & 22:49 &  59902.451 & 4160 &${\bf  1.192 }$&{\bf 0.067 }&${\bf  0.538 }$&{\bf 0.071 }&${\bf  0.407 }$&{\bf 0.050 }\\
                      & 2023-11-16 & 20:57 &  60265.373 & 3960 &${\bf  1.299 }$&{\bf 0.060 }&${\bf  0.646 }$&{\bf 0.063 }&${\bf  0.370 }$&{\bf 0.058 }\\ [1mm]  
{\rm WD\,2228$+$226  }& 2022-07-05 & 02:27 &  59765.602 & 3600 &${\rm -0.033 }$&{\rm 0.067 }&${\rm -0.056 }$&{\rm 0.046 }&${\rm  0.064 }$&{\rm 0.045 }\\ 
                      & 2022-11-19 & 21:34 &  59903.399 & 5200 &${\rm -0.046 }$&{\rm 0.060 }&${\rm -0.093 }$&{\rm 0.082 }&${\rm -0.035 }$&{\rm 0.076 }\\ [1mm]
{\rm WD\,2303$+$391  }& 2022-11-20 & 21:54 &  59904.412 & 6240 &${\rm -0.054 }$&{\rm 0.076 }&${\rm -0.229 }$&{\rm 0.105 }&${\rm -0.044 }$&{\rm 0.074 }\\ [1mm]
{\rm WD\,2345$+$027  }& 2022-11-19 & 23:06 &  59903.463 & 3960 &${\rm  0.082 }$&{\rm 0.077 }&${\rm -0.022 }$&{\rm 0.082 }&${\rm  0.142 }$&{\rm 0.090 }\\ [1mm]
{\rm WD\,2359$+$636  }& 2022-11-20 & 00:37 &  59903.526 & 4800 &${\rm -0.110 }$&{\rm 0.086 }&${\rm -0.120 }$&{\rm 0.075 }&${\rm -0.152 }$&{\rm 0.063 }\\ [1mm]

\hline
\end{longtable}
\end{onecolumn}

\end{center}
%%%%}
%%%%%%%%%%%%%%%%%%%%%
%\end{appendix}
\end{document}